\documentclass[aps,prmaterials,twocolumn,10pt,superscriptaddress]{revtex4-2}
\usepackage{amssymb,amsfonts,amsmath}
\usepackage{times}
\usepackage{graphicx}
\usepackage{color}
\usepackage{float}
\usepackage[T1]{fontenc}
\usepackage{cancel}
\usepackage[version=4]{mhchem}
\newlength{\figurewidth}
\newlength{\pagewidth}
\usepackage[a-1b]{pdfx}
\newcommand{\tc}{$T_\textrm{c}$}

\begin{document}

\title{Why Mg$_2$IrH$_6$ is predicted to be a high temperature superconductor, but Ca$_2$IrH$_6$ is not}

\author{Xiaoyu Wang}\affiliation{Department of Chemistry, State University of New York at Buffalo, Buffalo, USA}
\author{Warren Pickett}\affiliation{Department of Physics, University of California Davis, Davis, California, USA}
\author{Micheal Hutcheon}\affiliation{Deep Science Fund, Intellectual Ventures, Bellevue, Washington, USA}
\author{Rohit Prasankumar}\affiliation{Deep Science Fund, Intellectual Ventures, Bellevue, Washington, USA}
\author{Eva Zurek}\email{ezurek@buffalo.edu}\affiliation{Department of Chemistry, State University of New York at Buffalo, Buffalo, USA}

\begin{abstract}
The \ce{X2MH6} family, consisting of an electropositive cation X and a main group metal M octahedrally coordinated by hydrogen, has been predicted to hold promise for high-temperature conventional superconductivity. Herein, we analyze the electronic structure of two members of this family, \ce{Mg2IrH6} and \ce{Ca2IrH6}, showing why the former may possess superconducting properties rivaling those of the cuprates, whereas the latter does not. Within \ce{Mg2IrH6} the vibrations of the \ce{IrH6^4-} anions are key for the superconducting mechanism, and they induce coupling in the $e_g^*$ set, which are antibonding between the H 1$s$ and the Ir $d_{x^2-y^2}$ or $d_{z^2}$ orbitals. Because calcium possesses low-lying d-orbitals, $e_g^* \rightarrow$ \ce{Ca} $d$ back-donation is preferred, quenching the superconductivity. Our analysis explains why high critical temperatures were only predicted for second or third row {X} metal atoms, and may hold implications for superconductivity in other systems where the antibonding anionic states are filled.
\end{abstract}
\maketitle

\newpage

The theory-guided quest for hydride-based high-temperature conventional superconductors~\cite{Zurek:2021k} has led to the prediction of numerous systems belonging to the class of superhydrides ($e.g.$\ CaH$_4$\cite{Wang:2012,Zurek:2018b} and MgH$_4$\cite{Abe2018_PRB}, CaH$_6$~\cite{Wang:2012} and MgH$_6$\cite{Feng2015_RSCAdv}, Li$_2$CaH$_{16}$\cite{Zurek:2024f} and Li$_2$MgH$_{16}$\cite{RN754}). Some of these binary, simple systems, stable only at high pressures, have been synthesized and their record breaking superconducting critical temperatures, \tc s, have been measured ($e.g.$\ CaH$_6$ with a reported \tc\ of 210-215~K at 160-172~GPa~\cite{Ma2022_PRL,Li2022_NatCommun}). These successes have emboldened theoreticians to turn towards complex 1-atmosphere-metastable compounds, which are based upon hydrogen-containing molecular-building blocks, as the next set of candidates to be explored \emph{en route} towards the ultimate goal of ambient-condition superconductivity.

A number of recent studies have used workflows that coupled high-throughput Density Functional Theory (DFT) calculations and machine-learning techniques to screen a large number of compounds as potential high-\tc\ candidates, followed by high-fidelity electronic structure calculations and superconducting property predictions on the most promising systems~\cite{Choudhary_npj,Cerqueira2024_AdvMater,Cerqueira2024_AdvFunctMater,Sanna2024_npj,Zheng2024_MaterTodayPhys,Dolui2024_PRL}. From these studies, a new hope has emerged as being particularly fruitful for superconductivity, based on the cubic  \ce{X2MH6} family, where the cation X typically represents a group 1, 2, or 13 main group metal, and M is a transition (often a noble) metal that is octahedrally coordinated by hydrogen \cite{Dolui2024_PRL,Sanna2024_npj,Zheng2024_MaterTodayPhys,Cerqueira2024_AdvFunctMater}. Among this family, \ce{Mg2IrH6} has been predicted to exhibit a $T_\mathrm{c}$ potentially surpasses the boiling temperature of liquid nitrogen with estimates ranging from 65~K~\cite{Zheng2024_MaterTodayPhys} to 77~K~\cite{Sanna2024_npj}, and even as high as 160~K~\cite{Dolui2024_PRL}. Other systems belonging to this family with predicted \tc s ranging from $\sim$35-80~K include \ce{Mg2RhH6}, \ce{Mg2PdH6}, \ce{Mg2PtH6}~\cite{Sanna2024_npj}, \ce{Al2MnH6}, \ce{Li2CuH6} \cite{Zheng2024_MaterTodayPhys}, \ce{Na2AgH6}, \ce{Al2TcH6}, and \ce{Al2ReH6},\cite{Cerqueira2024_AdvFunctMater}.  Notably, numerous isotypic non-metallic compounds have been synthesized including \ce{Mg2RuH6}, \ce{NaCaIrH6} and \ce{Ca2IrH5}~\cite{Kadir2011_InorgChem}, and a pathway towards \ce{Mg2IrH6} via a recently synthesized \ce{Mg2IrH5} phase has been proposed~\cite{Strobel2024_arXiV}.

Though these studies have computed and listed the descriptors typically associated with superconductivity, including the electron-phonon-coupling (EPC) parameter, $\lambda$, logarithmic average frequency, $\omega_\text{log}$, density of states (DOS) at the Fermi level ($E_\text{F}$), $g(E_\text{F})$, and the superconducting gap, $\Delta_\textbf{k}$,  they have not analyzed the chemical features that result in high predicted \tc s for some of these compounds, but not for others. Herein, we perform such an analysis on the electronic structure of \ce{Mg2IrH6}, which has the largest computed \tc\ from this family of structures, and an isotypic \ce{Ca2IrH6} compound, which was reported as being non-superconducting~\cite{Sanna2024_npj}. Though Mg and Ca are both group 2 elements, low-lying unoccupied $d$-orbitals are present in Ca but not in Mg. Our analysis shows that back-donation from the \ce{IrH6^4-} anion to the Ca $d$ bands is the reason why superconductivity is quenched for \ce{Ca2IrH6}, and we discuss the implications on superconductivity in other systems in light of these findings.

\begin{figure*}[!ht]
    \centering
    \includegraphics[width=\textwidth]{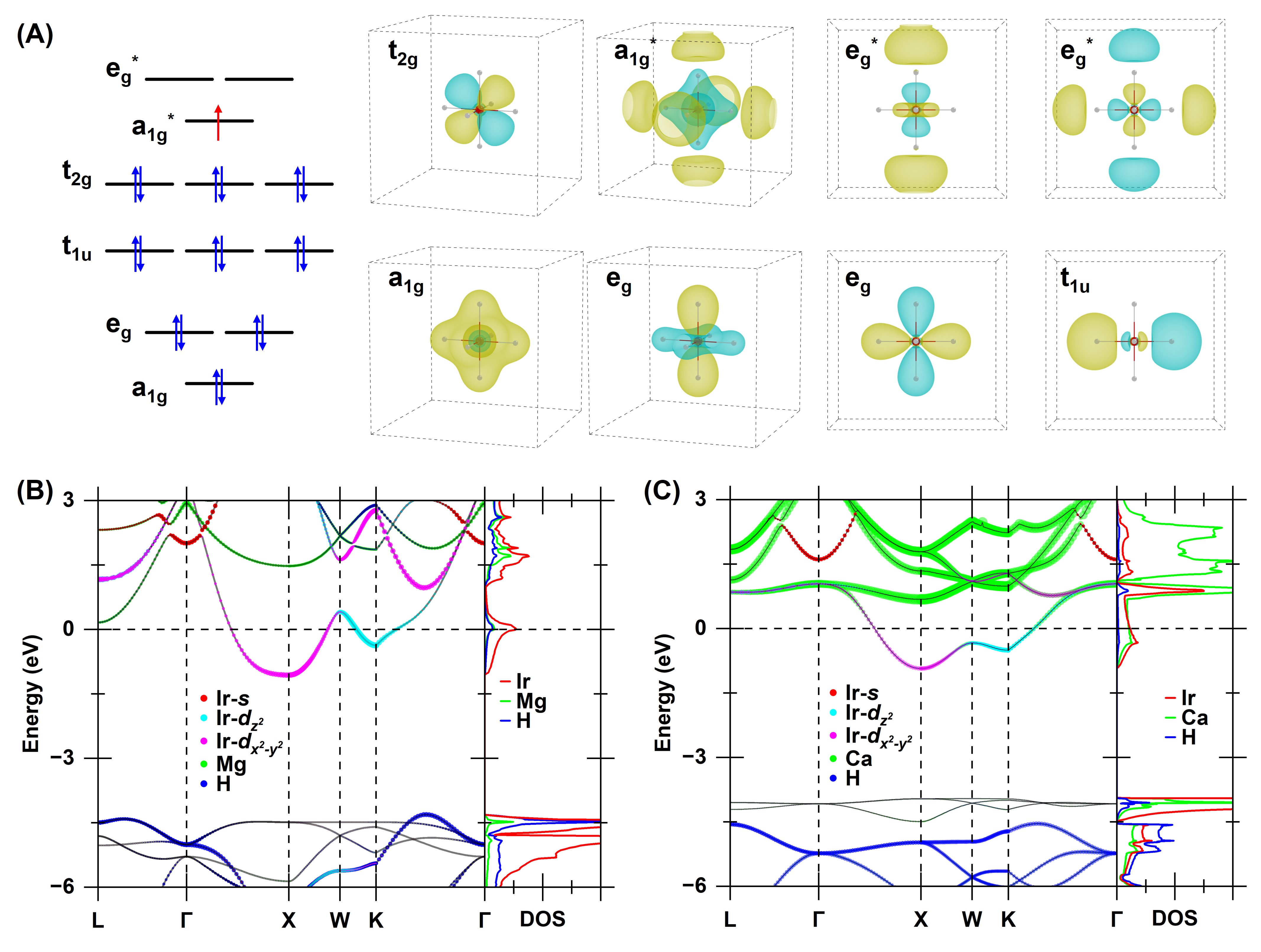}
    \caption{(A) Approximate molecular orbital (MO) diagram of \ce{IrH6^3-} anion (occupation shown denoted by the blue arrows), and isosurfaces of the resulting orbitals. The isovalue was chosen so that 35\% of the charge is encompassed within the surface. The red arrow illustrates the electron occupation for the \ce{IrH6^4-} anion. Bandstructure and density of states of (B) \ce{Mg2IrH6} and (C) \ce{Ca2IrH6}. The bandstructure is colored by projections onto specific \ce{Ir} orbitals and the Mg/Ca/H atoms, whereas the density of states is given as atomic projections.}
    \label{fig:electronic}
\end{figure*}

We begin, as every chemist would, by considering the molecular orbitals (MOs) of the closed-shell \ce{IrH6^3-} anion (Figure \ref{fig:electronic}A). Because the core states of \ce{Mg} ($2s^2,2p^6$) are too low in energy, and the empty states ($3s,3p$) are too high, the frontier MOs of this anion in the solid should contribute to the states close to $E_\text{F}$, and the band that crosses it. The highest occupied MO (HOMO) in \ce{IrH6^3-} turns out to be triply degenerate -- the $d_{xy}, d_{yz}, d_{xz}$ orbitals that point between the ligands, and do not interact with them, in accordance with predictions from ligand field theory (LFT). While LFT would suggest that the doubly degenerate $e_g^*$ orbitals, composed from an out-of-phase combination of the $d_{x^2-y^2}$ or $d_{z^2}$ orbitals with the hydrogen $s$ orbitals would correspond to the lowest unoccupied MOs (LUMOs), in our molecular calculations the $a_{1g}^*$, an antibonding combination of the Ir $6s$ and H $1s$ orbitals, falls between the $t_{2g}$ and $e_g^*$ sets. Both of the  $e_g^*$ orbitals turn out to have substantial contributions on the hydrogens, and are very diffuse pointing away from the Ir-H contact owing to the antibonding interaction. To this important point, we will return below.

In \ce{Mg2IrH6} the anion is \ce{IrH6^4-}. In this molecule the excess electron would occupy the $a_{1g}^*$ orbital, as shown in the schematic level diagram in Figure \ref{fig:electronic}A.  We relaxed the structure of the \ce{IrH6^4-} molecule with this electronic configuration, and an \ce{Ir}-\ce{H} bond length of 1.805~\r{A} was obtained, longer than the distance within the \ce{IrH6^3-} molecule (1.741~\r{A}) because the bonds are stretched when an electron fills the antibonding state. However, the Ir-H bond in the \ce{Mg2IrH6} crystal is unusual, 1.724~\r{A}, which is much shorter than in molecular \ce{IrH6^4-}, and closer to the value found in \ce{IrH6^3-}. To understand why this is so, we need to account for the periodicity of the structure, and inspect the Bloch functions formed by the symmetry-adapted linear combinations of the \ce{IrH6^4-} MOs~\cite{Hoffmann1987_Angew}. 

Though we cannot completely project the electronic bandstructure onto the MOs, we can project it onto those atomic orbitals (AOs) that contribute significantly to them. For example, projecting onto the Ir $6s$ should capture bands comprised of the $a_{1g}$ and the $a_{1g}^*$ MOs. The occupied $a_{1g}$ state is far below the energy scale that we plot in Figure~\ref{fig:electronic}B, but the $a_{1g}^*$, which possesses $s$-like symmetry, has its minimum energy, $\sim$2.5~eV above $E_\text{F}$, at the $\Gamma$ point where it interacts in-phase with  $a_{1g}^*$ orbitals in neighboring unit cells. The next set of MOs, the $e_g^*$, yield two bands that can be projected onto the $d_{z^2}$ and $d_{x^2+y^2}$ AOs of the \ce{Ir} atom. Along the $\Gamma$-$L$ high-symmetry line these two bands are degenerate and slightly higher in energy than a band with \ce{Mg} $s$-character. Along the $\Gamma$-$X$-$W$ path, however, the $d_{z^2}$ branch disperses below $E_\text{F}$, while the $d_{x^2-y^2}$ possesses a lower energy along the $\Gamma$-$K$-$W$ path. Therefore, unlike in molecular \ce{IrH6^4-}, where the unpaired electron fills the $a_{1g}^*$ orbital, within the \ce{Mg2IrH6} crystal the excess electron fills the $e_g^*$-based Bloch states, explaining the aforementioned discrepancy in the bond lengths. Within PBE we find the \ce{Mg2IrH6} crystal relaxes to a non-magnetic ground state with a remarkably high $g(E_\text{F})$, in-line with previous studies~\cite{Dolui2024_PRL,Sanna2024_npj}. For this geometry the ferromagnetic configuration is $\sim$25.4~meV/atom higher in energy.

\begin{figure*}[!ht]
    \centering
    \includegraphics[width=\textwidth]{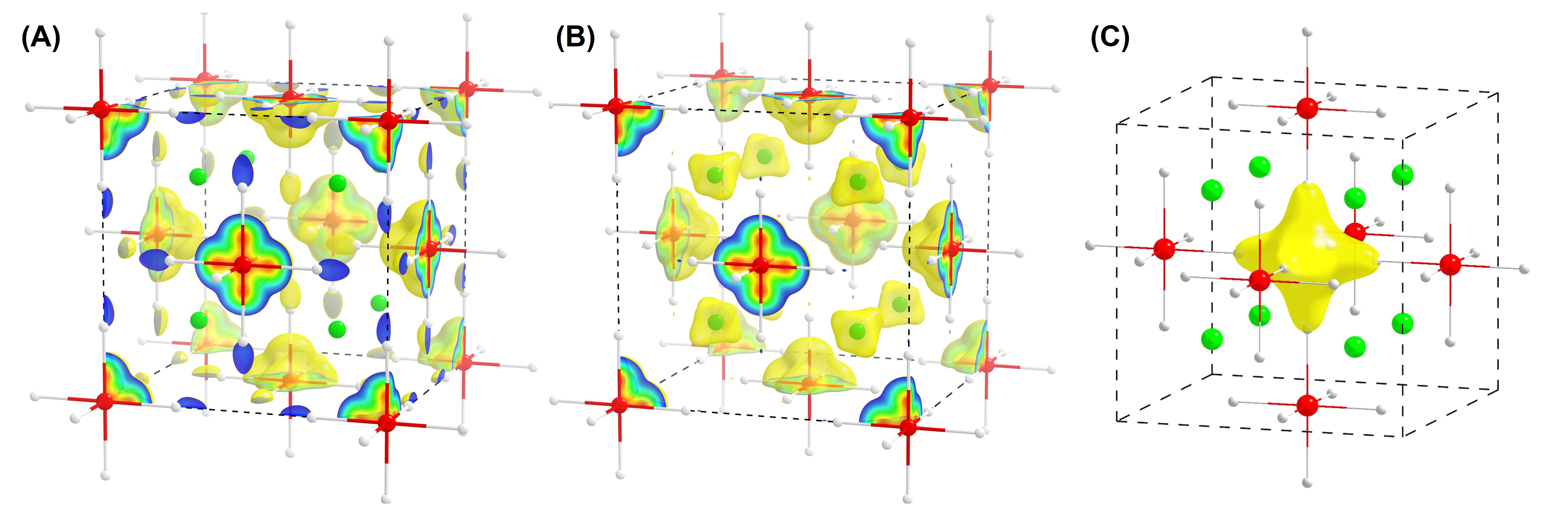}
    \caption{Charge density of the occupied metallic band calculated by integrating the charge within the energy window of $E_\mathrm{F}-2$~eV to $E_\mathrm{F}$ for (A) \ce{Mg2IrH6} and (B) \ce{Ca2IrH6}. Color code for atoms: Mg/Ca-green, Ir-red, and H-white. Isovalue was chosen as in Figure \ref{fig:electronic}. Color scale for contours are from blue (low density) to red (high density). (C) Isosurface of the charge density of the occupied metallic band computed on a grid that only contained points corresponding to the octahedral hole. The isovalue was chosen so that 50\% of the charge is encapsulated within the surface.}
    \label{fig:density}
\end{figure*}

Let us now turn to the isotypic \ce{Ca2IrH6}, which, somewhat surprisingly was not predicted to exhibit any superconductivity  ($T_\mathrm{c} \ll 0.1$~K)~\cite{Sanna2024_npj}. In this previous study, the radically different superconducting behavior of  \ce{Ca2IrH6} was attributed to the expansion of the lattice constant ($\sim$8\% larger than in \ce{Mg2IrH6}) that both weakens the interactions between neighboring \ce{IrH6^4-} units (resulting in a  smaller $g(E_\text{F})$), and leads to less disperse optical phonon modes, both which were proposed to result in a smaller $\lambda$~\cite{Sanna2024_npj}. To test this hypothesis, we calculated the \tc\ of \ce{Ca2IrH6} at 25~GPa -- a pressure where its lattice constant is almost identical to that of \ce{Mg2IrH6} at 1~atm -- but still we couldn't find superconductivity (Figure~S4). We therefore concluded that the absence of superconductivity in \ce{Ca2IrH6} cannot be explained by its lattice constant.

Comparison of the bandstructure calculated for \ce{Ca2IrH6} with that of \ce{Mg2IrH6} reveals a number of similarities, such as the way in which the metallic band ``runs''~\cite{Hoffmann1987_Angew}, but also some important differences (\emph{c.f.} Figure \ref{fig:electronic}B and C). Most notably, the metallic band contains significant \ce{Ca} $d_{xy}$, $d_{xz}$, and $d_{yz}$ character throughout, suggesting that these orbitals hybridize with the \ce{Ir} $d_{z^2}$ and $d_{x^2+y^2}$ states (Figure S2 and S3), lowering the energy of the band at the $W$-point so it drops below $E_\text{F}$. Such hybridization of the metal-cation and the anionic states is not observed in \ce{Mg2IrH6}. At first glance, this behavior is counter-intuitive: \ce{Ca} has smaller ionization potential (6.1~eV) than \ce{Mg} (7.6~eV), suggesting that it should have a greater proclivity to transfer its valence electron to \ce{IrH6^3-}, contrary to what we observe.

Digging deeper, we can understand why this happens nonetheless. The  $a_{1g}^*$ and $e_g^*$ orbitals within \ce{IrH6^3-} are antibonding and it is energetically unfavorable to fill them, but in \ce{Mg2IrH6} there is no other choice. However, \ce{Ca} contains low-lying $d$-orbitals available for \ce{IrH6^4-}$\rightarrow$~\ce{Ca} $d$ back-donation, which decreases the energy of the system. In fact, such back-donation, from H$_2$ $\sigma^*$ orbitals to Ca $d$, has been noted before in theoretical studies of high-pressure hydrides including CaH$_4$~\cite{Zurek:2018b,Zurek:2020a} and CaSH$_n$~\cite{Zurek:2020g}. The back-donation to the  \ce{Ca}-$d$ states in \ce{Ca2IrH6} depletes the charge density around $E_\mathrm{F}$ (integration of the electron density within an energy window of $E_\mathrm{F}\pm0.05$~eV yielded 0.26$e$/FU (formula unit) and 0.14$e$/FU  for \ce{Mg2IrH6} and \ce{Ca2IrH6}, respectively), thereby significantly decreasing $g(E_\text{F})$, which sits near the top of a peak in \ce{Mg2IrH6}, but is reduced by a factor of $\sim$2 within \ce{Ca2IrH6}.

\begin{figure*}[!ht]
    \centering
    \includegraphics[width=\textwidth]{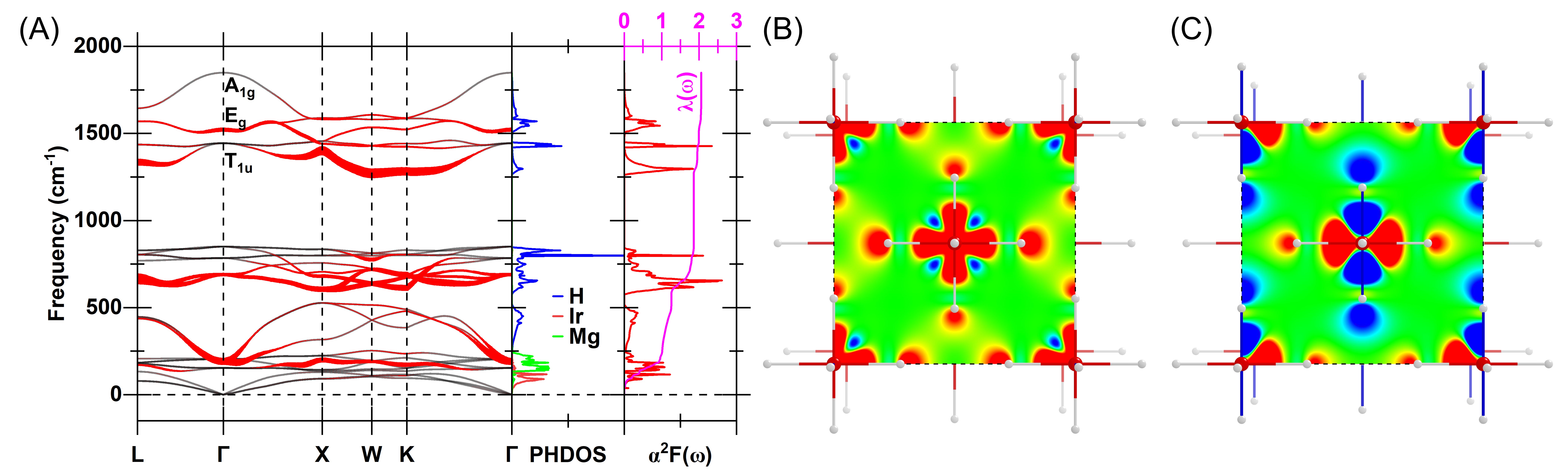}
    \caption{(A) Phonon bandstructure, projected phonon density of states, Eliashberg spectral function, $\alpha^2\mathrm{F}(\omega)$, and integral of the electron-phonon-coupling, $\lambda(\omega)$, for \ce{Mg2IrH6}. The thickness of the line is proportional to the projected $\alpha^2F(\omega)$. (B) Difference in the partial charge density for the occupied part of the metallic band between $E_g$-phonon-modulated structure and the unperturbed structure in the view of the $xy$ plane and (C) $yz$ plane. Charge gain/loss are represented with red/blue colors, while unchanged area is in green.}
    \label{fig:epc}
\end{figure*}

To investigate this further, we plotted the charge density of the occupied part of the metallic band for both systems (Figure \ref{fig:density}A and B). The topology of the \ce{IrH6^3-} $e_g^*$ orbitals are evident in the plot, notably the diffuse lobes on the hydrogen atoms that are directed towards the center of an octahedron, visualized in Figure \ref{fig:density}C. This overlap of diffuse antibonding orbitals that have a substantial contribution from hydrogen is reminiscent of the H$\leftrightsquigarrow$H bonding previously noted in metal-ammonia systems~\cite{Zurek:2009b}. In \ce{Ca2IrH6}, on the other hand, the lobes emanating from the hydrogen atoms are absent, and a tetrahedron of electron density surrounding each \ce{Ca} atom is evident instead. This tetrahedron is consistent with the projection of the charge density onto the  \ce{Ca} $t_{2}^*$ AOs, and the placement of the metal atom within a tetrahedral hole where four neighboring \ce{IrH6^4-} molecules reside at the tetrahedral vertices and induce a crystal field that splits the degeneracy of the \ce{Ca} $3d$ orbitals, which are not available for \ce{Mg}. At the $\Gamma$ point the $t_{2}^*$ orbitals are higher in energy than the $e^*$ set (Figure S3), as expected from the standard two-below-three splitting observed in molecular complexes. However, in other parts of the Brillouin Zone (BZ) this energy ordering reverses, as was discussed above for the anion-centered $a_g^*$ and $e_g^*$ based bands. To ensure that the computed charge density for the \ce{M2IrH6} crystals is not a result of the self-interaction or delocalization error present within non-hybrid DFT, we recalculated the electronic structure with the range-separated hybrid HSE06 functional (Figure~S1). The shape of the DOS near $E_\text{F}$ were virtually identical (for both metal atoms), strengthening our interpretation.

Now we are ready to analyze the coupling of the Bloch functions to the lattice vibrations: the EPC behavior. We decorated the phonon dispersion curve by a red line whose thickness is proportional to the value of the Eliashberg spectral function, $\alpha^2F(\omega)$, projected onto the phonon modes (calculated via scaling the phonon linewidth by the phonon frequency as $\gamma_{\mathbf{q}\nu}/\omega_{\mathbf{q}\nu}$) at any given wavevector $\mathbf{q}$ and phonon branch $\nu$ (Figure~\ref{fig:epc}A). A large EPC at a given frequency and for a particular vibrational mode should be accompanied by a large degree of valence electron redistribution. To illustrate this point, let us focus on the phonon bands above 1250~cm$^{-1}$ at the $\Gamma$ point, because their motions can be easily described in terms of vibrations that are well known for octahedral molecules. From these, only the $E_g$ mode (1520~cm$^{-1}$), which is related to the classic Jahn-Teller distortion of the \ce{IrH6^4-} octahedron, results in significant EPC. A single-point calculation was performed where the atoms were displaced along the eigenvectors corresponding to this mode with a displacement norm of 1~\r{A}, resulting in the elongation of two pairs of \ce{Ir}-\ce{H} bonds to $\sim$1.8~\r{A}, and the shrinking of the other pair to $\sim$1.5~\r{A}. The charge density difference between the phonon-modulated and the unperturbed structures ($\rho(\mathrm{modulated})-\rho(\mathrm{unperturbed})$) (Figure~\ref{fig:epc}B,C) shows a concomitant redistribution of electron density with a depletion of the population of the $d_{z^2}$, accompanied by an increase of the $d_{x^2-y^2}$. 

In contrast, neither the $A_{1g}$ (1850~cm$^{-1}$) nor the $T_{1u}$ (1444~cm$^{-1}$) modes induced any substantial EPC at $\Gamma$. The highest frequency mode is the symmetric stretch of \ce{IrH6^4-}, which maintains its $O_h$ symmetry, while the asymmetric $T_{1u}$ stretch modifies the $d_{z^2}$ and $d_{x^2-y^2}$ simultaneously. These vibrations do not break the degeneracy between $d_{z^2}$ and $d_{x^2-y^2}$, and therefore do not induce EPC around the $\Gamma$-point, though they have non-negligible contributions in other parts of the BZ. Notably, the vibrations from adjacent \ce{IrH6^4-} units can modify the amount of H$\leftrightsquigarrow$H overlap in the octahedral holes (Figure~\ref{fig:density}C), whose unperturbed charge density integrates to 0.15$e$ (as estimated by placing a sphere at the center with a radius of 1.14~\r{A}) from a total of 1$e$ in the occupied part of the metallic band. This yields EPC throughout the BZ, and not just at $\Gamma$. The decrease in the overall $g(E_\text{F})$, the nearly equal contribution of the Ca and \ce{IrH6^4-} based states to those near the Fermi level, and the heavier mass of \ce{Ca} as compared to \ce{Mg} are all factors that lead to a vanishingly small $T_\text{c}$ in \ce{Ca2IrH6} (0.024~K in our estimate).

By comparing and contrasting the electronic structures of two isotypic structures, \ce{Mg2IrH6} and \ce{Ca2IrH6}, we show why the former is predicted to be a high-$T_\text{c}$ superconductor, whereas the latter has a \tc\ near 0~K. Electron donation from the electropositive metals to the antibonding molecular orbitals in \ce{IrH6^3-} is energetically unfavorable, and if possible, back-donation to the unoccupied low-lying metal $d$-orbitals is preferred instead. This back-donation is impossible for metal atoms in the second or third row of the periodic table, but can readily occur for group 1, 2 or 13 main group metals in the fourth or higher rows. Since the EPC is associated with vibrations of the molecular anion, this charge reorganization is detrimental for superconductivity. Our findings can also explain previous results of high-throughput screening~\cite{Sanna2024_npj,Zheng2024_MaterTodayPhys} where high $T_\mathrm{c}$ was only discovered when the second and the third row metals served as the electron donors (\ce{Mg2RhH6}, \ce{Mg2IrH6}, \ce{Mg2PtH6}, \ce{Mg2PdH6}, \ce{Al2MnH6}, and \ce{Li2CuH6}).

Finally, back-donation to low-lying $d$-orbitals of alkaline earth metals from \ce{H2^\delta-} $\sigma^*$~\cite{Zurek:2020a} or \ce{CH4^\delta-} $a_1^*$~\cite{Zurek:2023m} antibonding orbitals may explain why compressed \ce{CaH4}, as well as \ce{CaC2H8} and \ce{SrC2H8} are not predicted to be good superconductors, whereas \ce{MgH4} and \ce{MgC2H8} have non-negligible \tc s within their range of dynamic stability. Back-donation to the metal $d$ bands would not be expected when the electropositive element donates electrons to bonding or non-bonding states, for example in \ce{CaB2} and \ce{MgB2} where the electron fills the $\pi$-bonding states instead \cite{Choi2009_PRB}.


\section*{Acknowledgements}
This work was supported by the Deep Science Fund of Intellectual Ventures and NSF award DMR-2136038. Calculations were performed at the Center for Computational Research at SUNY Buffalo (http://hdl.handle.net/10477/79221). We thank Stefano Racioppi for fruitful discussion.

%

%

\setlength{\bibsep}{0.0cm}
\bibliographystyle{Wiley-chemistry}
\bibliography{reference}

\clearpage



\end{document}